\documentclass[aps,prb,twocolumn,groupedaddress,showpacs]{revtex4-1}
\usepackage{graphics}
\usepackage{graphicx}
\usepackage{amsmath}
\usepackage{amssymb}
\usepackage{amsthm}
\usepackage{epsfig}
\usepackage{dsfont}
\begin{document}
\title{Comparative study of the Meissner and skin effects in superconductors}
\author{Jacob Szeftel$^1$}
\email[corresponding author :\quad]{jszeftel@lpqm.ens-cachan.fr}
\author{Nicolas Sandeau$^2$}
\author{Antoine Khater$^3$}
\affiliation{$^1$ENS Cachan, LPQM, 61 avenue du Pr\'esident Wilson, 94230 Cachan, France}
\affiliation{$^2$Aix Marseille Univ, CNRS, Centrale Marseille, Institut Fresnel, F-13013 Marseille, France}
\affiliation{$^3$Universit\'e du Maine, UMR 6087 Laboratoire PEC, F-72000 Le Mans France}
\begin{abstract}
The Meissner effect is studied by using an approach based on Newton and Maxwell's equations. The objective is to assess the relevance of London's equation and shed light on the connection between the Meissner and skin effects. The properties of a superconducting cylinder, cooled in a magnetic field, are accounted for within the same framework. The radial Hall effect is predicted. The energy, associated with the Meissner effect, is calculated and compared with the binding energy of the superconducting phase with respect to the normal one.  
\end{abstract}
\maketitle
		\section{introduction}
	The Meissner effect\cite{mei} highlights the rapid decay\cite{ash,gen,sch} of an applied magnetic field in bulk superconducting matter, provided the field is lower than some critical field. Our current understanding is still based mainly on London's equation\cite{lon}
\begin{equation}
\label{lon}
B+\mu_0\lambda^2_L\textrm{curl} j=0\quad,\quad\lambda_L=\sqrt{\frac{m}{\mu_0\rho e^2}}\quad,
\end{equation}
where $\mu_0,j,\lambda_L$ stand for the magnetic permeability of vacuum, the persistent current, induced by the magnetic induction $B$ and London's length, whereas $e,\quad m,\quad\rho$ refer to the charge, effective mass and concentration of superconducting electrons, respectively. Eq.(\ref{lon}), combined with the Amp\`ere-Maxwell equation, entails\cite{lon} that the penetration depth of the magnetic field is equal to $\lambda_L$. The validity of Eq.(\ref{lon}) was questioned early by skin depth measurements\cite{pip1} (see also\cite{par} p.37, $2^{nd}$ paragraph, lines $7,8$).\par
	Furthermore, because the Meissner-Ochsenfeld experiment\cite{mei} yielded merely qualitative results, it was widely believed, till London's work\cite{lon}, that $H$ did not penetrate \textit{at all} into the superconducting sample. Thus, since the Meissner-Ochsenfeld experiment failed to provide an accurate\cite{gen} assignment for the field penetration length, all of the experiments\cite{pip1,h1,gor} have consisted of measuring the penetration depth of an electromagnetic wave into a superconductor, i.e. the skin depth\cite{jac,bor}, at frequencies in the range $\left[10MHz,100GHz\right]$. The penetration of the electromagnetic field into a conductor is impeded by the real part of the frequency dependent dielectric constant, associated with conduction electrons, being negative below the plasma frequency. The skin effect has been analyzed\cite{sze} previously in superconductors and found to have essentially the same properties, as those observed in a normal conductor.\par
	 The purpose of this work is then to show theoretically that the properties of the Meissner effect are conditioned  by the limited time-duration $t_0$, needed in the experiment, for the applied magnetic field $H$ to grow with time $t$ from its starting value $H(t=0)=0$ up to its permanent one $H(t_0)$. The analysis relies entirely on Newton and Maxwell's equations.  In particular, it will appear below that the spatial decay of the \textit{static} field $H(t> t_0)$ inside the bulk superconductor, characterizing the Meissner effect, can be described as a sum involving $\delta(n\omega_0),n=1,2,3...$, where $\delta(\omega)$ is the frequency dependent skin depth and $\omega_0=\frac{2\pi}{t_0}$.\par
	  The outline is as follows : Sections II and III deal with the skin effect in superconductors and with the Meissner effect, respectively, while establishing the connection between both effects. The validity of Eq.(\ref{lon}) is assessed in Section IV. The case of the field cooled superconductor is addressed in Section V. The radial Hall effect is analyzed in Section VI. The energy, associated with the Meissner effect, is calculated in Section VII. The conclusions are given in Section VIII.\par
	Consider as in Fig.1 a superconducting material of cylindrical shape, characterized by its symmetry axis $z$ and radius $r_0$ in a cylindrical frame with coordinates ($r,\theta,z$). The superconducting sample is further inserted into a coil, producing $H(t)$. Both lengths of the superconducting sample and of the coil are taken $>>r_0$, in order to get rid of any end effect, which will turn out to be a crucial requirement for the Hall effect experiment. The superconducting  material contains electrons of charge $e$, effective mass $m$, and concentration $\rho$. The current $I(t)$, flowing through the coil, gives rise, thanks to the Faraday-Maxwell law, to an electric field $E_\theta(t,r)$, normal to the unit vectors along the $r$ and $z$ coordinates, such that $E_\theta(t,r)\neq 0$ for $t\in ]0,t_0[$ only, which defines a transient regime $(0<t<t_0\Leftrightarrow E_\theta\neq 0)$ and a permanent one $(t>t_0\Leftrightarrow E_\theta= 0)$. 
		\section{transient regime}
$E_\theta$ induces a current $j_\theta(t,r)$ along the field direction, as given by Newton's law
\begin{equation}
\label{newt}
\frac{dj_\theta}{dt}=\frac{\rho e^2}{m}E_\theta-\frac{j_\theta}{\tau}\quad,
\end{equation}
where $\frac{\rho e^2}{m}E_\theta$ and $-\frac{j_\theta}{\tau}$ are respectively proportional to the driving force accelerating the conduction electrons and a friction term. The friction force $\propto \frac{j_\theta}{\tau}$ in Eq.(\ref{newt}) ensues from the \textit{finite} conductivity, observed in superconductors, carrying an \textit{ac current} (see\cite{sch} p.4, $2^{nd}$ paragraph, lines $9,10$). For example, for the superconducting phase of $BaFe_2(As_{1-x}P_x)_2$, the conductivity, measured in the microwave range, has been found (see\cite{h1} p.1555, $3^{rd}$ column, $2^{nd}$ paragraph, line $11$) to be $\approx 300\sigma_n$, where $\sigma_n$ stands for the normal conductivity, measured just above the critical temperature $T_c$. Additional evidence is provided by commercial microwave cavity resonators, made up of superconducting materials, displaying a very high, albeit \textit{finite} conductivity. At last note that the finite conductivity, measured at $\omega\neq 0$ in superconductors, is consistent with the observation of persistent currents at \textit{vanishing} electric field, even though a cogent explanation is still lacking\cite{ash,gen,sch}.\par 
\begin{figure}
\includegraphics*[height=6 cm,width=8 cm]{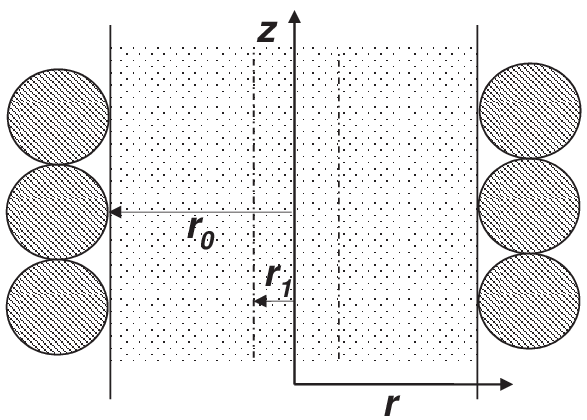}
\caption{Cross-section of the superconducting sample (dotted) and the coil (hatched); $E_\theta,j_\theta$ are both normal to the unit vectors along the $r$ and $z$ coordinates, whereas $B_z,H_z$ and $E_r,j_r$ are parallel to the unit vector along the $z$ and $r$ coordinates, respectively; the radius of the wire making up the coil has been magnified for the reader's convenience; the matter between the dashed lines should be carved out to carry out the radial Hall effect experiment}
\label{Bzr}
\end{figure}
	$E_\theta$ induces a magnetic induction $B_z(r,t)$, parallel to the $z$ axis. $B_z$ is given\cite{sze} by the Faraday-Maxwell equation as
\begin{equation}
\label{Bz}
-\frac{\partial B_z}{\partial t}=\frac{E_\theta}{r}+\frac{\partial E_\theta}{\partial r}\quad,
\end{equation}
while the magnetic field $H_z(t,r)$, parallel to the $z$ axis, is given\cite{sze} by the Amp\`ere-Maxwell equation as
\begin{equation}
\label{Hz}
-\frac{\partial H_z}{\partial r}=2j_\theta+\epsilon_0\frac{\partial E_\theta}{\partial t} \quad.
\end{equation}
 $\epsilon_0$ refers to the electric permittivity of vacuum.\par
	$E_\theta\left(t,r\right),j_\theta\left(t,r\right),B_z\left(t,r\right),H_z\left(t,r\right)$ can be recast as Fourier series for $t\in ]0,t_0[$
\begin{equation}
\label{fou}
\begin{array}{l}
f\left(t,r\right)=\sum_{n\in\mathds{Z}} f\left(n,r\right)e^{in\omega_0 t}\quad,\\ f\left(n,r\right)=\int_0^{t_0}e^{-in\omega_0 t}f\left(t,r\right)dt/t_0\quad,
\end{array}
\end{equation}
where $\omega_0 t_0=2\pi$ and $f\left(t,r\right),f\left(n,r\right)$ hold for  $B_z\left(t,r\right)$, $H_z\left(t,r\right)$, $E_\theta\left(t,r\right)$, $j_\theta\left(t,r\right)$ and $B_z\left(n,r\right)$, $H_z\left(n,r\right)$, $E_\theta\left(n,r\right)$, $j_\theta\left(n,r\right)$, respectively. Replacing $E_\theta,j_\theta,B_z,H_z$ in Eqs.(\ref{newt},\ref{Bz},\ref{Hz}) by their expression in Eqs.(\ref{fou}), while taking into account
$$B_z\left(n,r\right)=\mu\left(n\omega_0\right)H_z\left(n,r\right)\quad,$$
where $\mu\left(n\omega_0\right)=\mu_0\left(1+\chi_s\left(n\omega_0\right)\right)$ and $\chi_s\left(\omega\right)$ is the magnetic susceptibility of superconducting electrons at frequency $\omega$, yields for $n\neq 0$
\begin{equation}\label{fou2}
\begin{array}{l}
E_\theta\left(n,r\right)=\frac{1+in\omega_0\tau}{\sigma}j_\theta\left(n,r\right)\\
in\omega_0 B_z\left(n,r\right)=-\left(\frac{E_\theta\left(n,r\right)}{r}+\frac{\partial E_\theta\left(n,r\right)}{\partial r}\right)\\
\frac{\partial B_z\left(n,r\right)}{\partial r}=-\mu\left(n\omega_0\right)\left(2j_\theta\left(n,r\right)+in\omega_0\epsilon_0E_\theta\left(n,r\right)\right)
\end{array}
\end{equation}
where the conductivity $\sigma=\frac{\rho e^2\tau}{m}$\cite{ash}. Then eliminating $E_\theta\left(n,r\right),j_\theta\left(n,r\right)$ from Eqs.(\ref{fou2}) gives
\begin{equation}
	\label{skin}
\frac{\partial^2 B_z\left(n,r\right)}{\partial r^2}=\frac{B_z\left(n,r\right)}{\delta^2(n\omega_0)}-\frac{\partial B_z\left(n,r\right)}{r\partial r}\quad.
\end{equation}
$$\delta(\omega)=\frac{\lambda_L}{\sqrt{\left(1+\chi_s\left(\omega\right)\right)\left(\frac{2i\omega\tau}{1+i\omega\tau}-\frac{\omega^2}{\omega^2_p}\right)}}\quad,\quad\omega_p=\sqrt{\frac{\rho e^2}{\epsilon_0m}}$$
refer to skin depth and plasma frequency\cite{ash,jac,bor}, respectively. The solution of Eq.(\ref{skin}) with $\frac{dB_z}{dr}\left(r=0\right)=0$, is a Bessel function, such that $B_z(r>>|\delta(n\omega_0)|)\approx e^{r/\delta(n\omega_0)}$.\par
		\section{permanent regime}
Because $E_\theta(t>t_0,r)=0$ in the permanent regime, the friction force $\propto-\frac{j_\theta}{\tau}$ is no longer at work for $t>t_0$, so that the \textit{transient} current $j_\theta(t< t_0)$ turns into the \textit{persistent} one, $j_\theta(t> t_0,r)=j_\theta(t_0,r),\forall r$. Eqs.(\ref{fou}) then yield
\begin{equation}
\label{jr}
j_\theta(t>t_0,r)=2\sum_{n\in\mathds{Z}, n\neq0} j_\theta\left(n,r\right)\quad.
\end{equation}
  The Amp\`ere-Maxwell equation reads now
\begin{equation}
\label{max2}
-\frac{\partial H_z}{\partial r}(t> t_0,r)=j_\theta(t_0,r) \quad.
\end{equation}
Comparing Eqs.(\ref{Hz},\ref{max2}) reveals that $H_z(t_{0^-},r)\neq H_z(t_{0^+},r)$. The penetration depth $\lambda_M$ of the static field $H_z(t_{0^+},r)$ is defined as
$$\frac{1}{\lambda_M}=\frac{\partial \textrm{Log}H_z(t_{0^+},r_0)}{\partial r}\quad .$$
  Because of $r_0>>|\delta(n\omega_0)|$ under typical experimental conditions, there is $j_\theta(n,r\rightarrow r_0)\approx j_\theta(n,r_0)e^{\frac{r-r_0}{\delta(n\omega_0)}}$. Using Eq.(\ref{jr}) to integrate Eq.(\ref{max2}), we obtain
\begin{equation}
\label{mei2}
\frac{1}{\lambda_M}\approx\frac{\sum_n  j_\theta(n,r_0)}{\sum_n\delta(n\omega_0)j_\theta(n,r_0)} \quad,
\end{equation}
where the sum is performed for $n\neq 0$ and $|n|\omega_0<\omega_p$. Eq.(\ref{mei2}) embodies the connection between the skin and Meissner effects. Due to $|\delta(n\omega_0)|=\lambda_L/\sqrt{2|n|\omega_0\tau}>>\lambda_L$ for $|n|\omega_0\tau<<1$, it ensues from Eq.(\ref{mei2}) that $|\lambda_M|>>\lambda_L$. Furthermore, by contrast with Eq.(\ref{lon}), it is obvious from the $\omega_0$ dependence in Eq.(\ref{skin}) that there can be no one to one correspondence between $H_z(t>t_0,r)$ and $j_\theta(t>t_0,r)$, which is an irreversible consequence of the friction term $-\frac{j_\theta}{\tau}$ in Eq.(\ref{newt}). This is anyhow of little practical interest because $\lambda_M$ cannot be measured, as already noted in Section I.
	\section{validity of London's equation}
	Eq.(\ref{lon}) was assumed\cite{lon}, starting from the following version of Newton's equation
\begin{equation}
\label{newt1}
\frac{dj_\theta}{dt}=\frac{\rho e^2}{m}E_\theta\quad,
\end{equation}
which is identical to Eq.(\ref{newt}) in case $\tau \rightarrow\infty$. Integrating both sides of Eq.(\ref{newt1}) from $t=0$ up to $t=t_0$ yields for for $r\in [0,r_0]$
\begin{equation}
\label{lon1}
j_\theta(t_0,r)=\frac{\rho e^2}{m}\int_0^{t_0}E_\theta(t,r)dt=-\frac{\rho e^2}{m}A_\theta(t_0,r)\quad,
\end{equation}
by assuming $j_\theta(t=0,r)=A_\theta(t=0,r)=0$ and taking advantage of $E_\theta=-\frac{\partial A_\theta}{\partial t}$, where the magnetic vector potential\cite{jac,bor} $A_\theta(t,r)$ is parallel to $E_\theta$. Using furthermore $B_z=\textrm{curl} A_\theta$, it is inferred from Eq.(\ref{lon1}) for $r\in [0,r_0]$ in the permanent regime $t>t_0$
$$B_z+\mu_0\lambda^2_L\textrm{curl} j_\theta=0\quad,$$
which is identical to Eq.(\ref{lon}). The validity of London's equation has thence been shown provided $\tau\rightarrow\infty$. 
	\section{field cooled sample}
	As the susceptibility $\chi_s$ not being continuous at $T_c$ ($T_c$ refers to the critical temperature) will turn out to be \textit{solely} responsible for the Meissner effect to occur in a superconductor, cooled inside a magnetic field, we set out to reckon it. Since no paramagnetic contribution is observed in the superconducting state\cite{ash,gen,sch}, the latter is deemed to be in a macroscopic singlet spin state, so that the only contribution to $\chi_s$ can be calculated using Maxwell's equations. We begin with writing down the $t$-averaged density of kinetic energy
$$\mathcal{E}_K(r)=\frac{m}{2\rho}\left(\frac{j_\theta(r)}{e}\right)^2\quad,$$
associated with the \textit{ac} current $j_\theta(r)e^{i\omega t}$, flowing along the $E_\theta$ direction (this latter induces in turn a magnetic field $H_z(r)e^{i\omega t}$, parallel to the $z$ axis). The Amp\`ere-Maxwell equation simplifies into $\frac{\partial H_z}{\partial r}=-2j_\theta$ for practical $\omega<<\omega_p$. As this discussion is limited to the case $r\rightarrow r_0$, both $H_z(r),j_\theta(r)$ are $\propto e^{r/\delta(\omega)}$, so that $\mathcal{E}_K(r)$ is recast into
\begin{equation}
\label{k2}
\mathcal{E}_K(r)=\frac{\mu_0}{8}\left(\frac{\lambda_L}{|\delta(\omega)|}H_z(r)\right)^2\quad.
\end{equation}
Moreover there is the identity $\frac{\partial \mathcal{E}_K}{\partial M}=-H_z$, where $M=\mu_0\chi_s(\omega)H_z$ is the magnetization of superconducting electrons. Actually this identity reads in general $\frac{\partial F}{\partial M}=-H_z$, where $F$ represents the Helmholz free energy\cite{lan}. However the property that a superconducting state carries no entropy\cite{ash,gen,sch} entails that $F=\mathcal{E}_K$. Equating this expression of $\frac{\partial \mathcal{E}_K}{\partial M}$ with that inferred from Eq.(\ref{k2}) yields finally
$$\chi_s(\omega)=-\left(\frac{\lambda_L}{2|\delta(\omega)|}\right)^2\quad.$$
As expected, $\chi_s$ is found diamagnetic $(\chi_s<0)$ and $|\chi_s(\omega)|<<1$ for $\omega<<1/\tau$. The calculation of $\chi_s(0)$ proceeds along the same lines, except for the Amp\`ere-Maxwell equation reading $\frac{\partial H_z}{\partial r}=-j_\theta$ and $\lambda_M$ showing up instead of $\delta(\omega)$, whence
$$\chi_s(0)=-\left(\frac{\lambda_L}{\lambda_M}\right)^2\quad.$$
 Note that our definition of $\chi_s=\frac{M(r)}{\mu_0H_z(r)}$, where $H_z(r),M(r)$ refer to local field and magnetization at $r$, differs from the usual\cite{ash,lon,gen} one $\chi_s=\frac{M}{\mu_0H_z(r_0)}$ with $H_z(r_0),M$ being external field and total magnetization.\par
 While the sample is in its normal state at $T>T_c$, the applied magnetic field $H_z$ penetrates fully into bulk matter and induces a magnetic induction
$$B_n=\mu_0\left(1+\chi_n\right)H_z\quad,$$
where $\chi_n$ designates the magnetic susceptibility of conduction electrons. It comprises\cite{ash} the sum of a paramagnetic (Pauli) component and a diamagnetic (Landau) one and $\chi_n>0$ in general. Moreover the magnetic induction reads for $T<T_c(H_z)$
$$B_s=\mu_0\left(1+\chi_s(0)\right)H_z\quad,$$
with $\chi_s(0)<0$. Because of $\chi_s(0)\neq\chi_n$, the magnetic induction undergoes a finite step while crossing $T_c(H_z)$
\begin{equation}
\label{dBt}
\frac{\delta B}{\delta t}=\frac{B_s-B_n}{\delta t}=\mu_0\frac{\chi_s(0)-\chi_n}{\delta t}H_z\quad,
\end{equation}
where $\delta t$ refers to the time needed in the experimental procedure for $T$ to cross $T_c(H_z)$. Due to the Faraday-Maxwell equation (see Eq.(\ref{Bz})), the finite $\delta B/\delta t$ induces an electric field $E_\theta$ such that
$\textrm{curl} E_\theta=-\frac{\delta B}{\delta t}$, giving rise to the persistent, $H_z$ screening current.\par
	Noteworthy is that, though $H_z$ remains \textit{unaltered} during the cooling process, the magnetic induction $B$ is indeed \textit{modified} at $T_c$, as shown by Eq.(\ref{dBt}). This $B$ variation arouses the driving force, giving rise to the screening current $j_\theta$, and ultimately to $H_z$ expulsion, as explained in Section II, III.\par
	\section{the radial Hall effect}
	For $t<t_0$, the magnetic induction $B_z$ exerts on the conduction electrons a radial Lorentz force $\frac{B_zj_\theta}{\rho}$, pushing the electrons inward, so that a charge distribution builds up, which in turn gives rise, via Poisson's law, to a radial electric field $E_r(r)$, characterizing the Hall effect. Meanwhile $E_r$ drives a transient radial current $j_r(t)$ ($j_r(t<t_0)\neq 0$, $j_r(t>t_0)=0$), responsible for the charge distribution.\par
	 Moreover for $t>t_0$, equilibrium is secured by the radial centrifugal force $\frac{m}{r}\left(\frac{j_\theta(t_0,r)}{\rho e}\right)^2$, exerted on each electron making up the persistent current $j_\theta(t_0,r)$, being counterbalanced by the sum of the Lorentz force and an electrostatic one $eE_r$, with $E_r$ given by
\begin{equation}
\label{Erad}
E_r=-\frac{j_\theta}{\rho e}\left(B_z+\frac{m}{\rho e^2r}j_\theta\right)\quad.
	\end{equation}
Owing to the Amp\`ere-Maxwell equation $j_\theta=-\frac{\partial H_z}{\partial r}$ and $B_z=\mu_0H_z$, $E_r$ can be recast as
$$E_r=\frac{\mu_0}{\rho e}\frac{\partial H_z}{\partial r}\left(H_z-\frac{\lambda_L^2}{r}\frac{\partial H_z}{\partial r}\right)\quad.$$
 Because of $\frac{\partial H_z}{\partial r}\approx \frac{H_z}{\lambda_M}$, $r_0>>\lambda_L$ and $|\lambda_M|>>\lambda_L$, the approximation $E_r\approx \frac{\mu_0}{2\rho e}\frac{\partial H_z^2}{\partial r}$ can be used for significant $r>>\lambda_L$. For the Hall effect to be observed, a sample in shape of a hollow cylinder of inner and outer radius $r_1,r_0$, respectively, is needed (see Fig.\ref{Bzr}). Furthermore the length of the sample should be larger than that of the coil and the measurement should be carried out in the middle of the sample to get rid of any end effect. Finally the Hall voltage reads, for $r_0-r_1>>|\lambda_M|$
$$U_H=-\int_{r_1}^{r_0}E_r(r)dr\approx -\frac{\mu_0}{2\rho e}H_z^2\left(t_0^+,r_0\right)\quad.$$
  As in normal metals\cite{ash}, measuring $U_H$ gives access to $\rho$. However, whereas $H_z,j_\theta$ are set independently from each other in the Hall effect observed in a normal conductor, which implies $U_H\propto H_zj_\theta$, they are both related by the Amp\`ere-Maxwell equation in the experiment discussed hereabove, which entails $U_H\propto H_z^2$. Note also that $U_H$ is independent of $r_1$ provided $r_0-r_1>>|\lambda_M|$. As $\lambda_M$ cannot be measured, the Hall effect is likely to provide the only way to assess the validity of this work.\par
 	\section{calculation of the energy}
	The whole energy, associated with the Meissner effect, comprises two contributions, i.e. the kinetic energy, carried by the persistent current, and the electrostatic one, stemming from the Hall effect. Taking advantage of the the Amp\`ere-Maxwell equation, the density of kinetic energy is inferred to read
\begin{equation}
\label{k3}
\mathcal{E}_K(r)=\frac{m}{2\rho}\frac{j^2_\theta(t_0,r)}{e^2}=\frac{\mu_0}{2}\left(\frac{\lambda_L}{\left|\lambda_M\right|}H_z(t_0^-,r)\right)^2\quad,
\end{equation}
where $j_\theta(t_0,r)$ refers to the persistent current and $\frac{\lambda_L}{\left|\lambda_M\right|}<<1$.\par
	The expression of the radial current $j_r$ is needed to reckon the electrostatic energy. To that end we look for a solution of the Amp\`ere-Maxwell equation with no magnetic field. Thus it reads :
$$j_r+\frac{\partial D_r}{\partial t}=2j_r+\epsilon_0\frac{\partial E_r}{\partial t}=0\quad,$$
where the electric displacement\cite{sze} $D_r$ is parallel to the unit vector, along the $r$ coordinate, and the time derivative of the space charge, stemming from the Hall effect, has been taken to vanish, so that $\frac{\partial P_r}{\partial t}=j_r\Rightarrow\frac{\partial D_r}{\partial t}=j_r+\epsilon_0\frac{\partial E_r}{\partial t}$, with $P_r$ being the radial polarisation\cite{sze}. This assumption is vindicated by the sought electrostatic energy, depending \textit{only} on the permanent electric field $E_r(t_0,r)$, and accordingly being \textit{independent} from the preceding transient behaviour $E_r(t<t_0,r)$. Multiplying then the above equation by $E_r$ yields
$$E_rj_r+\frac{\epsilon_0}{4}\frac{\partial E_r^2}{\partial t}=0\quad.$$
Using Eq.(\ref{Erad}), we obtain
$$j_r\left(\frac{j_\theta}{\rho e}B_z\right)dt=-E_rj_rdt=\frac{\epsilon_0}{4}\frac{\partial E_r^2}{\partial t}dt\quad.$$
 The left hand term is identified as the elementary work performed by the Lorentz force, so that the searched expression of the the density of electrostatic energy is obtained, thanks to the first law of thermodynamics, as
$$\mathcal{E}_e(r)=\frac{\epsilon_0}{4}\int_0^{t_0}\frac{\partial E_r^2}{\partial t}dt=\frac{\epsilon_0}{4}E_r^2(t_0,r)=\mu_0\frac{H_z^4(t_0,r)}{\left(2c\rho e|\lambda_M|\right)^2}\quad,$$
where $c$ stands for the light velocity in vacuum. By replacing $H_z$ by its upper bound $H_c$, it can be checked that $\mathcal{E}_e<<\mathcal{E}_K$ in all cases.\par
 	In the mainstream treatment\cite{gen,sch,lon}, the energy, pertaining to the Meissner effect, has rather been conjectured to read $\mathcal{E}_M=\frac{\mu_0}{2}H_z^2(r)>>\mathcal{E}_K(H_z)$ in Eq.(\ref{k3}), due to $\lambda_L<<|\lambda_M|$. Moreover this expression of $\mathcal{E}_M$ turns out to be questionable from another standpoint, because its value for $H_z=H_c(T)$ is furthermore believed \cite{ash,gen,sch,lon} to be equal to $\rho E_b(T)$, where $2E_b(T)$ designates the binding energy, needed to turn a pair of BCS electrons into two normal ones, at temperature $T$. However the BCS theory\cite{bar} provides the estimate $\frac{E_b}{E_F}\approx \left(\frac{k_BT_c}{E_F}\right)^2$, where $E_F,k_B$ stand for the Fermi energy in the normal state and Boltzmann's constant, respectively. A numerical application with $E_b=\mathcal{E}_M(H_c(0))/\rho$ in the case of $Al$ yields $T_c\approx 10^{-5}K$, i.e. much less than the measured value $T_c=1.19K$.\par
	 Likewise multiplication of both terms of Eq.(\ref{newt}) by $j_\theta$ and time-integration yield the following inequality
$$\frac{m}{\rho e^2}\int_0^{t_0}j_\theta\frac{dj_\theta}{dt}dt=\mathcal{E}_K(H_c(T))<<\int_0^{t_0}j_\theta(t) E_\theta(t) dt\quad,$$
with $j_\theta(t) E_\theta(t)$ being the external power fed into the sample at $t$. Actually it ensues from Ohm's law, recast as
 $$\frac{\sigma}{\tau}E_\theta=\frac{\rho e^2}{m}E_\theta=\frac{j_\theta}{\tau}\quad,$$ 
because the inertial force, $\propto |\frac{dj_\theta}{dt}|$ in Eq.(\ref{newt}), is negligible\cite{sze} with respect to the electric one $\propto\frac{\rho e^2}{m}|E_\theta|$, provided $\left|\frac{dj_\theta}{dt}\frac{\tau}{j_\theta}\right|<<1$, which always holds for the Meissner-Ochsenfeld experiment. 
	\section{conclusion}
	The applied, time-dependent magnetic field excites transient eddy currents according to Newton and Maxwell's equations, which turn to persistent ones, after the magnetic field stops varying and the induced electric field thereby vanishes. Those eddy currents thwart the magnetic field penetration. Were the same experiment to be carried out in a normal metal, eddy currents would have built up the same way. However, once the electric field vanishes, they would have been destroyed quickly by Joule dissipation and the magnetic field would have subsequently penetrated into bulk matter. As a matter of fact, the Meissner effect shows up as a \textit{classical} phenomenon and a mere outcome of \textit{persistent currents}, the very signature of superconductivity. The common physical significance of the Meissner and skin effects has been unveiled too. The radial Hall effect has been predicted. The energy, associated with the Meissner effect, has been calculated and compared with the binding energy of the superconducting phase. 

\end{document}